%% file: main.tex
\documentclass[conference, 10pt]{IEEEtran}
\IEEEoverridecommandlockouts
\usepackage{cite}
\usepackage{amsmath,amssymb,amsfonts}
\usepackage{mathtools}
\usepackage{graphicx}
\usepackage{subcaption}
\usepackage{textcomp}
\usepackage{xcolor}
\usepackage{tabularx}
\usepackage{booktabs}
\usepackage{float}
\usepackage{algpseudocode}
\usepackage[english]{babel}
\usepackage{paralist}
\usepackage{array}
\usepackage{amsmath,stackrel}
\usepackage{multicol}
\usepackage{verbatim}
\usepackage{enumerate}              
\usepackage{dsfont}
\usepackage{psfrag}
\usepackage{algorithm}
\usepackage[nolist]{acronym}
\usepackage[utf8]{inputenc}
\usepackage{tikz}
\usepackage{collcell}
\usepackage{pgfplots}
\pgfplotsset{compat=1.18}
\usepackage{hhline}
\usepackage{colortbl}
\usepackage{multido}
\usepackage{multirow}
\usepackage{dblfloatfix}    
\usepackage{etoolbox}
\usepackage{soul}
\usepackage{eucal}
\usepackage{cuted}
\usepackage[font=footnotesize]{caption}

\newcommand{\Jm}{\mathbf{J}_\mathrm{\Omega}}
\newcommand{\Jinvm}{\mathbf{J}_{\mathrm{\Omega}^{-1}}}

\newcommand{\transp}{\mathrm{T}}

\makeatletter
\let\oldfootnote\footnote
\def\footnote{\@ifstar\footnote@star\footnote@nostar}
\def\footnote@star#1{{\let\thefootnote\relax\footnotetext{#1}}}
\def\footnote@nostar{\oldfootnote}
\makeatother
\input{acronyms.tex}

\def\BibTeX{{\rm B\kern-.05em{\sc i\kern-.025em b}\kern-.08em
    T\kern-.1667em\lower.7ex\hbox{E}\kern-.125emX}}

\DeclareMathOperator{\Tr}{Tr}
\begin{document}


\title{Bistatic Sensing at THz Frequencies via a Two-Stage Ultra-Wideband MIMO-OFDM System
\thanks{This work was supported in part by the European Union under the Italian National Recovery and Resilience Plan (NRRP) of NextGenerationEU, partnership on ``Telecommunications of the Future'' (PE00000001 - program ``RESTART''), and in part by the HORIZON-JU-SNS-2022 project TIMES (Grant Agreement Number: 101096307).}
}
\author{\IEEEauthorblockN{
Tommaso~Bacchielli, 
Lorenzo~Pucci,
Davide~Dardari, 
and Andrea~Giorgetti
}
\medskip
\IEEEauthorblockA{Wireless Communications Laboratory, CNIT, DEI, University of Bologna, Italy\\ 
Email: \{tommaso.bacchielli2, lorenzo.pucci3, davide.dardari, andrea.giorgetti\}@unibo.it\\
}

}

\maketitle

\begin{abstract}
The availability of abundant bandwidth at \ac{THz} frequencies holds promise for significantly enhancing the sensing performance of \ac{ISAC} systems in the next-generation wireless systems, enabling high accuracy and resolution for precise target localization. In \ac{OFDM} systems, wide bandwidth can be achieved by increasing the subcarrier spacing rather than the number of subcarriers, thereby keeping the complexity of the sensing system low. However, this approach may lead to an ambiguity problem in target range estimation. To address this issue, this work proposes a two-stage maximum likelihood method for estimating target position in an ultra-wideband bistatic multiple-antenna \ac{OFDM}-based \ac{ISAC} system operating at \ac{THz} frequencies. Numerical results show that the proposed estimation approach effectively resolves the ambiguity problem while achieving high resolution and accuracy target position estimation at a very low signal-to-noise ratio regime.
\end{abstract}
\vskip -0.2cm

\acresetall

\section{Introduction}
\Ac{ISAC} at \ac{THz} frequencies is a key promising enabler to meet requirements for next-generation wireless systems, where the ubiquitous integration of sensing, connectivity, and intelligence across various domains will be crucial \cite{Tong_Zhu_2021}. 
For an \ac{ISAC} system, the vast bandwidth availability at \ac{THz} ($0.1-10\,$THz) holds the potential of exploiting ultra-wide bandwidth, leading to the simultaneous transmission of a large amount of data with unprecedented data rates \cite{THz_channel}, alongside very high range estimation accuracy and resolution \cite{THz_survey}. 

Most of the studies regarding \ac{ISAC} have focused on \ac{OFDM}-based systems, primarily at sub-6 GHz and millimeter wave frequencies, exploring both monostatic and bistatic radar configurations \cite{Barneto2019,Braun,pucci2022system,Pucci_Bistatic}. However, little investigation has been conducted to exploit the potential of \ac{ISAC} system operating at \ac{THz} frequencies. In \cite{Beam_THz}, a novel \ac{ISAC} technique with joint beam-squint and beam-split for \ac{THz} massive \ac{MIMO} system is presented. The authors show the advantage of the beam-squint to simultaneously transmit towards various directions and the possibility of beam-splitting to extend sensing range and enhance accuracy using only a few subcarriers. In \cite{THz_ISAC}, the authors evaluate the range estimation performance of a system at \ac{THz} frequencies by analyzing different waveforms. This work demonstrates that \ac{OFDM} is still an excellent candidate for \ac{ISAC}, also at \ac{THz}, but focuses on systems with limited bandwidth, in the order of MHz. 

To the author's knowledge, there remains a lack of research investigating the potential advantages of utilizing \ac{OFDM}-based \ac{ISAC} systems at \ac{THz} with ultra-wide bandwidth.
Since the coherence bandwidth of short-range \ac{THz} channels is hundreds of MHz, it is possible to achieve very high bandwidth by properly setting the subcarrier spacing without a large amount of subcarriers. However, while this solution ensures controlling the complexity, it may result in a low maximum unambiguous range, which reduces the system's operating range \cite{Braun}.

In this context, this work proposes to mitigate the range estimation ambiguity problem inherent to large subcarrier spacings with a novel two-stage transmission scheme followed by a two-stage \ac{ML} estimation approach at the receiver.
This approach first performs coarse target estimation over a sub-band with a smaller subcarrier spacing to avoid the ambiguity problem, followed by refined super-resolution estimation using the whole system bandwidth. Numerical results demonstrate that the proposed system can localize the target without ambiguity in a confined indoor area, ensuring cm-level range resolution and mm-level accuracy.

In this paper, we adopt the following notation: capital boldface letters for matrices, lowercase bold letters for vectors, $\Tr(\cdot)$ for the trace of a  matrix, $(\cdot)^\ast$, $(\cdot)^\mathrm{T}$ and $(\cdot)^\mathrm{H}$ for conjugate, transpose and conjugate transpose of a vector/matrix, respectively, and $|\cdot|$, $||\cdot||_2$ for the modulus and Euclidean norm, respectively. Also, $\otimes$, $\mathbb{E}[\cdot]$, $\mathrm{var}[\cdot]$ for Kronecker product, statistical mean value and variance operator. $[\mathbf{X}]_{a:b,c:d}$ denotes a submatrix of $\mathbf{X}$ composed of a subset $[a,b]$ of rows and $[c,d]$ of columns.
A zero-mean circularly symmetric complex Gaussian random vector with covariance $\boldsymbol{\Sigma}$ is denoted by $\mathbf{x} \thicksim \mathcal{CN}( \mathbf{0},\boldsymbol{\Sigma})$, while $x \sim \mathcal{U}(a,b)$ represents a uniformly distributed random variable $x$ within the interval $(a,b)$. 

The rest of the paper is organized as follows. In Section~\ref{sec:system_model} and \ref{sec:two_stage}, the system model and the proposed two-stage \ac{ML} estimation method are introduced. Numerical results are presented in Section~\ref{sec:num_Res}, while Section~\ref{sec:conclusion} concludes the paper.


\section{System Model} \label{sec:system_model}

This work considers a bistatic \ac{MIMO} \ac{OFDM}-based \ac{ISAC} system operating at \ac{THz} frequencies. In particular, for what concerns sensing, such a system consists of a \ac{Tx} and a \ac{Rx} to form a bistatic pair. Both the \ac{Tx} and the \ac{Rx} are \acp{BS} used to communicate with \acp{UE} or machinery while simultaneously localizing targets in the surrounding environment by using downlink signals. Differently from an \ac{ISAC} monostatic system, in which a full-duplex architecture and self-interference cancellation techniques are required \cite{Barneto2019}, in the bistatic configuration these problems are naturally solved as \ac{Tx} and \ac{Rx} are not co-located \cite{Pucci_Bistatic}. However, in this case, the \ac{Tx} and the \ac{Rx} need to be synchronized to perform target localization, e.g., by connecting them to a central unit that coordinates sensing operations. This work assumes that \ac{Tx} and \ac{Rx} are synchronized.

Without loss of generality, both \ac{Tx} and \ac{Rx} are equipped with a \ac{ULA} composed of $N_\mathrm{t}$ and $N_\mathrm{r}$ antenna elements half-wavelength spaced, respectively.
To perform sensing tasks, an \ac{OFDM} frame consisting of $M$ \ac{OFDM} symbols with $K$ active subcarriers is transmitted. The $k$-th subcarrier has frequency $f_k=f_\mathrm{c}+k\Delta f$, with $k=-K/2,\dots,K/2-1$, where $f_\mathrm{c}$ is the carrier frequency and $\Delta f$ the subcarrier spacing. The total frame duration is equal to $MT_\mathrm{s}$, being $T_\mathrm{s} \triangleq 1/\Delta f+T_\mathrm{cp}$ the total \ac{OFDM} symbol duration including the \ac{CP}; the total bandwidth is therefore $B=K\Delta f$.
Moreover, to exploit the \ac{ULA}, beamforming is performed at the \ac{Tx} through a precoding vector, $\mathbf{w}$. 

\subsection{Input-Output Relationship} \label{sec:I/0}
The complex envelope of the continuous-time \ac{OFDM} signal transmitted by the \ac{Tx} antenna array is defined as
\begin{equation}
    \mathbf{s}(t) = \sqrt{\frac{P_\mathrm{t}G_\mathrm{t}}{K}}\mathbf{w} \sum_{m=0}^{M-1}\left( \sum_{k=0}^{K-1}x[k,m]e^{j2 \pi k \Delta f t}\right)g(t-m T_\mathrm{s})   
\end{equation}
where $P_\mathrm{t}$ is the total transmitted power, $G_\mathrm{t}$ is the single antenna element gain at the \ac{Tx}, and $\mathbf{w} \in \mathbb{C}^{N_\mathrm{t} \times 1}$ represents a unit-norm beamforming vector designed to illuminate with an almost constant gain a relatively wide angular sector where a target $p$ is expected to be located. Moreover, $x[k,m]$ denotes a generic complex modulation symbol transmitted at time instant $m$ on subcarrier $k$, while $g(t)$ represents the modulating pulse.

After the \ac{FFT} block in the \ac{OFDM} demodulator at the \ac{Rx} and taking into account negligible \ac{ICI} and \ac{ISI}, a received time-frequency grid of complex elements $y[k,m]$ is obtained at each antenna element. By aggregating the symbols at each antenna, a vector $\mathbf{y}[k,m] \in \mathbb{C}^{N_\mathrm{r} \times 1}$ of received complex modulation symbols is obtained, as follows
\begin{equation} \label{eq:Rx_signal}
    \mathbf{y}[k,m] = \sqrt{\frac{P_\mathrm{t}G_\mathrm{t}G_\mathrm{r}}{K}}\mathbf{H}_{\mathrm{t},\mathrm{r}}[k,m] \mathbf{w} x[k,m] +\mathbf{\tilde{n}}[k,m]
\end{equation}
where $G_\mathrm{r}$ is the single antenna element gain at the \ac{Rx}, $\mathbf{H}_{\mathrm{t},\mathrm{r}}[k,m] \in \mathbb{C}^{N_\mathrm{r} \times N_\mathrm{t}}$ is the \ac{MIMO} channel matrix between the \ac{Tx} and the \ac{Rx} for the subcarrier $k$ at time $m$, and $\mathbf{\tilde{n}} \sim \mathcal{CN}(\mathbf{0},\sigma_{\tilde{n}}^2\mathbf{I}_{N_\mathrm{r}})$ represents the \ac{AWGN}, with $\sigma_{\tilde{n}}^2= N_0 \Delta f$. Here, $N_0 = k_\mathrm{B} 290 n_\mathrm{F}$ is the noise power spectral density, with $k_\mathrm{B}$ the Boltzmann constant and $n_\mathrm{F}$ the receiver noise figure.

\subsection{Channel Model} \label{sec:channel_model}
Without loss of generality, this work considers a single point-like target scenario. Since a bistatic setting is considered, the channel matrix at subcarrier $k$ and time $m$ can be obtained as the outer product of the channel vector $\mathbf{h}_{p,\mathrm{r}}[k,m]\in \mathbb{C}^{N_\mathrm{r}\times1}$, which represents the channel between the target $p$ and the \ac{Rx}, with the vector $\mathbf{h}_{\mathrm{t},p}[k,m] \in \mathbb{C}^{1 \times N_\mathrm{t}}$ representing the channel between the \ac{Tx} and the target $p$. These vectors are given by
\begin{equation}
    \mathbf{h}_{p,\mathrm{r}}[k,m]=\beta_{p,\mathrm{r}}e^{j2\pi(mT_\mathrm{s}f_{\mathrm{D}_{p,\mathrm{r}}}-k\Delta f\tau_{p,\mathrm{r}})}\mathbf{b}(\theta_{p,\mathrm{r}})
    \label{eq:rx_channel}
\end{equation}
\begin{equation}
    \mathbf{h}_{\mathrm{t},p}[k,m]=\alpha_{\mathrm{t},p}e^{j2\pi(mT_\mathrm{s}f_{\mathrm{D}_{\mathrm{t},p}}-k\Delta f \tau_{\mathrm{t},p})}\mathbf{a}^\mathrm{H}(\phi_{\mathrm{t},p})
    \label{eq:tx_channel}
\end{equation}
respectively. In \eqref{eq:tx_channel}, $\alpha_{\mathrm{t},p} = \sqrt{\xi_{\mathrm{t},p}}e^{-j(2\pi f_\mathrm{c}\tau_{\mathrm{t},p}+\varphi_0)}$ represents the complex channel coefficient associated with target $p$, where $\xi_{\mathrm{t},p}$ denotes the channel gain, and $\varphi_0 \in \mathcal{U}{[0,2\pi)}$ accounts for the phase shift between \ac{Tx} and \ac{Rx}. Furthermore, $\tau_{\mathrm{t},p} = r_{\mathrm{t},p}/c$ represents the propagation delay with respect to target $p$, where $r_{\mathrm{t},p}$ is the distance between target $p$ and \ac{Tx}; $f_{\mathrm{D}_{\mathrm{t},p}}$ and $\phi_{\mathrm{t},p}$ denote the Doppler shift and \ac{AoD} associated with target $p$, respectively. The transmit array response, denoted by $\mathbf{a}(\phi_{\mathrm{t},p}) \in \mathbb{C}^{N_\mathrm{t} \times 1}$, is defined considering far-field propagation conditions with the center of the array as a reference, as follows
\begin{equation}
    \mathbf{a}(\phi_{\mathrm{t},p})=\left[e^{-j\frac{N_\mathrm{t}-1}{2}\pi \sin{(\phi_{\mathrm{t},p})}},\dots,e^{j\frac{N_\mathrm{t}-1}{2}\pi \sin{(\phi_{\mathrm{t},p})}}\right]^\mathrm{T}.
    \label{eq:steering_vec}
\end{equation}
When \ac{LoS} propagation conditions are in place, the path loss coefficient is given by
   $\xi_{\mathrm{t},p} = \frac{\sigma_{p}}{4\pi r^2_{\mathrm{t},p}}$,
where $\sigma_{p}$ is the \ac{RCS} of target $p$.

Similarly, in \eqref{eq:rx_channel}, $\tau_{p,\mathrm{r}}=\gamma_{p,\mathrm{r}}/c$ is the propagation delay from the target $p$ to the center of the \ac{Rx}, with $\gamma_{p,\mathrm{r}}$ the Euclidean distance between $p$ and the \ac{Rx}. Furthermore, $\beta_{p,\mathrm{r}}=\sqrt{\zeta_{p,\mathrm{r}}}e^{-j2\pi f_\mathrm{c}\tau_{p,\mathrm{r}}}$ is the channel coefficient, and $\mathbf{b}(\theta_{p,\mathrm{r}})~\in~\mathbb{C}^{N_\mathrm{r} \times 1}$ is the array response vector defined as in \eqref{eq:steering_vec}, being $\theta_{p,\mathrm{r}}$ the \ac{AoA} related to the target $p$.
Considering \ac{LoS} propagation conditions and isotropic antenna elements with effective area $A_\mathrm{eff}=c^2/(4\pi f_\mathrm{c}^2)$, the path loss factor is given by
    $\zeta_{p,\mathrm{r}}=\frac{c^2}{(4\pi f_\mathrm{c} \gamma_{p,\mathrm{r}})^2}$.

Starting from \eqref{eq:rx_channel} and \eqref{eq:tx_channel}, the $N_\mathrm{r} \times N_\mathrm{t}$ channel matrix in \eqref{eq:Rx_signal}, which represents the channel between \ac{Tx}  and \ac{Rx} for subcarrier $k$ at time $m$, can be modeled as
\begin{equation} \label{eqn:H}
    \mathbf{H}_{\mathrm{t},\mathrm{r}}[k,m] = \epsilon_p e^{j2\pi(mT_\mathrm{s}f_{\mathrm{D}_p}-k\Delta f\tau_p)} \mathbf{b}(\theta_{p,\mathrm{r}}) \mathbf{a}^{\mathrm{H}}(\phi_{\mathrm{t},p})
\end{equation}
where $\epsilon_p=\alpha_{\mathrm{t},p}\beta_{p,\mathrm{r}}$ and $\tau_p=\tau_{\mathrm{t},p}+\tau_{p,\mathrm{r}}$ are the bistatic complex channel factor and bistatic propagation delay, respectively, associated with the target $p$. The bistatic Doppler shift $f_{\mathrm{D}_p}$ is defined as
$f_{\mathrm{D}_p}=f_{\mathrm{D}_{\mathrm{t},p}}+f_{\mathrm{D}_{p,\mathrm{r}}}=\frac{2|\mathbf{v}_p|}{\lambda}\cos{(\delta)}\cos{(\beta/2)}$,
where $\mathbf{v}_p$ is the velocity of the target $p$, while $\beta$ and $\delta$ are the bistatic angle and the angle between the velocity direction and the bisector of the bistatic angle $\beta$, respectively \cite{Pucci_Bistatic}.

\subsection{Received Signal and Signal-to-Noise Ratio} \label{sec:rx_signal_snr}
By replacing \eqref{eqn:H} in \eqref{eq:Rx_signal} and stacking the $K \times M$ transmitted symbol grid into a $KM \times 1$ vector $\mathbf{\underline{x}}$, a more compact form of the vector of received symbols $\mathbf{\underline{y}} \in \mathbb{C}^{N_\mathrm{r}KM \times 1}$ can be obtained as follows
\begin{equation}
    \label{eq:Rx_signal_2}
    \mathbf{\underline{y}} = \sqrt{\frac{P_\mathrm{t}G_\mathrm{t}G_\mathrm{r}}{K}}\epsilon_p \mathbf{G}(f_{\mathrm{D}_p},\tau_p, \theta_{p,\mathrm{r}},\phi_{\mathrm{t},p})\mathbf{\underline{x}}+\mathbf{\tilde{\underline{n}}}
\end{equation}
where $\mathbf{G}(\cdot)$ is the effective channel matrix of dimension $N_\mathrm{r}KM \times KM$, defined for a generic target $p$, as
\begin{align} 
\label{eqn:G}
    \mathbf{G}(f_{\mathrm{D}_p},\tau_p, \theta_{p,\mathrm{r}},\phi_{\mathrm{t},p})
   \triangleq \mathbf{T}(\tau_p,f_{\mathrm{D}_p}) \otimes \left(\mathbf{b}(\theta_{p,\mathrm{r}}) \mathbf{a}^\mathrm{H}(\phi_{\mathrm{t},p}) \mathbf{w} \right)
\end{align}
with $\mathbf{T}(\tau_p,f_{\mathrm{D}_p}) \in \mathbb{C}^{KM \times KM}$ defined as
\begin{align}
    \mathbf{T}(\tau_p,f_{\mathrm{D}_p}) \triangleq \mathrm{diag} \Bigl([1,\dots,&e^{j2\pi (M-1)T_\mathrm{s}f_{\mathrm{D}_p}}]^\mathrm{T}\otimes\\
    &\otimes [1,\dots,e^{-j2\pi (K-1)\Delta f \tau_p}]^\mathrm{T} \Bigr). \nonumber
\end{align}

Recalling the complex channel factor $\epsilon_p$ and the noise variance $\sigma^2_{\tilde{n}}$ introduced in Section~\ref{sec:channel_model}, the \ac{SNR} of the bistatic radar system related to target $p$ can be defined as
\begin{align} \label{eq:SNR}
    \mathrm{SNR} & = \frac{\mathbb{E}\bigl\{\bigl|x[k,m]\bigr|^2\bigr\}P_\mathrm{t}G^a_\mathrm{t} G_\mathrm{r}|\epsilon_p|^2}{K\sigma^2_{\tilde{n}}}\\
    & = \mathbb{E}\bigl\{\bigl|x[k,m]\bigr|^2\bigr\}\frac{P_\mathrm{t}G^a_\mathrm{t} G_\mathrm{r}\sigma_{p} c^2}{(4\pi)^3 r^2_{\mathrm{t},p} \gamma^2_{p,\mathrm{r}} f^2_\mathrm{c}N_0 K \Delta f}. \nonumber
\end{align}
where $G^a_\mathrm{t}$ is the total \ac{Tx} antenna array gain in the direction of $p$, considering both $G_\mathrm{t}$ and beamforming gain.
If convenient, when performing numerical simulations varying the \ac{SNR}, \eqref{eq:SNR} 
can be recast as $\text{SNR} = 1/\sigma^2_{\tilde{n}}$, by normalizing the transmitted complex symbols to average unit transmit power as $\mathbb{E}\bigl\{\bigl|x[k,m]\bigr|^2\bigr\} = 1$ and setting $|\epsilon_p|^2=1$. 

\subsection{Target Position Estimation}
In this work, the bistatic propagation delay $\tau_p$, which is directly related to the bistatic range $r_p = \tau_p c$, and the \ac{AoA} $\theta_p$, are estimated via a \ac{ML} approach. 

As already mentioned, at \ac{Tx} we perform beamforming through a vector $\mathbf{w}$ designed to provide nearly constant gain over a wide circular sector. Therefore, at the \ac{Rx}, the array response from \ac{Tx} to the target is a complex constant, i.e., $g_p = \mathbf{a}^\mathrm{H}(\phi_p) \mathbf{w}$, which can be absorbed in the channel gain coefficient, as $h_p \triangleq g_p \epsilon_p$. The effective channel gain thus becomes independent on the \ac{AoD} $\phi_p$, i.e., $\mathbf{G}(f_{\mathrm{D}_p},\tau_p,\theta_p)$.

The vector of sensing parameters related to the target $p$ to be estimated is defined as $\boldsymbol{\Theta}_p~=~(|h_p|, \angle{h_p}, f_{\mathrm{D},p}, \tau_p, \theta_p)$, so the \ac{ML} estimate of the set $\boldsymbol{\Theta}_p$ involves a search in the space $\Gamma \triangleq \mathbb{C} \times \mathbb{R}^3$ \cite{dehkordi2023multistatic}
\begin{equation} \label{eqn:ML}
    \boldsymbol{\Theta}_{\mathrm{ML}} = \underset{\boldsymbol{\Theta}_p \in \Gamma}{\arg\min} \; \left\| \mathbf{\underline{y}} - \sqrt{\frac{P_\mathrm{t}G_\mathrm{t}G_\mathrm{r}}{K}}h_p \mathbf{G}(f_{\mathrm{D}_p},\tau_p,\theta_p) \mathbf{\underline{x}}\right\|^2_2.
\end{equation}

One possible solution of \eqref{eqn:ML} can be obtained through a \ac{GML} approach. It consists of estimating the sensing parameters of interest in $\boldsymbol{\Theta}_p$ by maximizing the log-likelihood function with respect to the channel gain coefficient $h$. This reduces the sensing parameters \ac{ML} estimation to the solution of the following problem
\begin{equation} \label{eqn:ML_param}
    \left(\hat{f}_{\mathrm{D}_p}, \hat{\tau}_p,\hat{\theta}_p\right) = \mathrm{arg} \; \underset{(f_\mathrm{D},\tau,\theta) \in \Psi}{\mathrm{max}} \; \frac{|\boldsymbol{\xi}(\tau, f_\mathrm{D}) \mathbf{b}^*(\theta)|^2}{\mathbf{\underline{x}}^\mathrm{H}\mathbf{b}^\mathrm{H}(\theta)\mathbf{b}(\theta)\mathbf{\underline{x}}}
\end{equation}
where $\Psi$ is a grid composed of a set of $(f_\mathrm{D},\tau,\theta)$ tuples, and $\boldsymbol{\xi}(\tau,f_\mathrm{D})=\left[\mathbf{\underline{x}}^\mathrm{H} \mathbf{T}^\mathrm{H}(\tau,f_\mathrm{D}) \mathbf{Y}^\mathrm{T}\right]$, with $\mathbf{Y}=\left[\mathbf{\underline{y}}_1[0,0],\dots,\mathbf{\underline{y}}_{N_\mathrm{r}}[M-1,K-1]\right] \in \mathbb{C}^{N_\mathrm{r} \times KM}$.
The position of the target $p$ is thus obtained from the estimate of the bistatic range $\hat{r}_p=\hat{\tau}_p c$ and \ac{AoA} $\hat{\theta}_p$, by first computing the distance from target $p$ to \ac{Rx}, as
\begin{equation}
    \hat{\gamma}_p = \frac{\hat{r}^2_p-L^2}{2\left(\hat{r}_p+L\sin{(\hat{\theta}_p-\pi/2)}\right)}
\end{equation}
where $L$ is the distance between \ac{Tx} and \ac{Rx}, commonly referred to as \emph{baseline}.
Then, the estimated position of the target with respect to the \ac{Rx} is  $\mathbf{\hat{p}}_p=[\hat{\gamma}_p\cos{(\hat{\theta}_p)},\hat{\gamma}_p\sin{(\hat{\theta}_p)}]^\mathrm{T}$.

\subsection{Position Error Bound} \label{sec:PEB}
Let us consider the received signal in \eqref{eq:Rx_signal_2}, and define $\mathbf{\underline{s}} =~ \sqrt{\frac{P_\mathrm{t}G_\mathrm{t}G_\mathrm{r}}{K}}h_p\, \mathbf{G}(f_{\mathrm{D}_p},\tau_p,\theta_p)\, \mathbf{\underline{x}} \in \mathbb{C}^{N_\mathrm{r} K M \times 1}$ as the vector of mean values of $\mathbf{\underline{y}}$. The \ac{CRLB} on the estimation of the unknown parameters in $\mathbf{\Theta}_p$ is
\begin{equation} \label{eqn:Fisher_var}
    \mathrm{var}[{\hat{\Theta}_{p_i}}] \geq  \mathrm{CRLB}(\hat{\Theta}_{p_i})=I_{p_{ii}}^{-1}\qquad i=1,\dots,\mathrm{card(\mathbf{\Theta}_p)}
\end{equation}
\noindent where ${\hat{\Theta}_{p_i}}$ is the estimate of the $i$-th parameter contained in $\boldsymbol{\Theta}_p$, and $I_{p_{ii}}^{-1}$ represents the $i$-th element on the main diagonal of the inverse of the \ac{FIM} $\mathbf{I}_p$. The generic $ij$-th element of $\mathbf{I}_p$ can be computed as \cite{Rife1974, BacPucPaoGio:C23}
\begin{equation} \label{eqn:Fisher}
    I_{p_{ij}} = \frac{2}{\sigma^2_{\tilde{n}}} \mathfrak{Re} \Biggl\{ \sum_{k,m,n} \biggl(\frac{\partial s_n[k,m]}{\partial \Theta_{p_i}}\biggr)^\ast \biggl(\frac{\partial s_n[k,m]}{\partial \Theta_{p_j}}\biggr) \Biggl\}
\end{equation}
where $s_n[k,m]$ is the $(n,k,m)$ element of a 3D array obtained starting from the $N_\mathrm{r} K M \times 1$ vector $\mathbf{\underline{s}}$ and given by 
\begin{equation} \label{eq:s_n}
    s_n[k,m] = \sqrt{\frac{P_\mathrm{t}G_\mathrm{t}G_\mathrm{r}}{K}}h_p e^{j2\pi (m T_\mathrm{s} f_{\mathrm{D}_p}-k \Delta f \tau_p)} b_n(\theta_p) x[k,m]
\end{equation}
being $b_n(\theta_p)$ the $n$-th element of the array response vector $\mathbf{b}$.
After computing the inverse of the \ac{FIM} $\mathbf{I}_p$, the bound on position estimation accuracy, hereinafter referred to as \ac{PEB}, is given by $\mathrm{PEB} = \sqrt{\mathrm{CRLB}(\mathbf{\hat{p}}_p)}$,
where the \ac{CRLB} of position estimate can be found as \cite{jourdan2008position}
\begin{align} \label{eq:CRLB_bis}
    \mathrm{CRLB}(\mathbf{\hat{p}}_p)&=\Tr(\Jm^{-1} [\mathbf{I}_p^{-1}(\boldsymbol{\Theta}_p)]_{4:5,4:5}[\Jm^{-1}]^\transp) \nonumber \\
    & =\Tr([\mathbf{I}_p^{-1}(\boldsymbol{\Theta})]_{4:5,4:5}[\Jinvm]^\transp\Jinvm)
\end{align}
being $\Jm$ the Jacobian of the transformation $\Omega:~\mathbf{\hat{p}}_p~=~(x_p,y_p) \rightarrow (\tau_p,\theta_p)$, and $\Jinvm$ the Jacobian of the inverse transformation $\Omega^{-1}:(\tau_p,\theta_p) \rightarrow \mathbf{\hat{p}}_p=(x_p,y_p)$, defined as\footnote{Note that the following property holds $\Jinvm = \Jm^{-1}$, and vice versa.}
\begin{equation} \label{eq:Jinvm}
    \Jinvm=
    \begin{bmatrix}
        \frac{\partial x_p}{\partial \tau_p} & \frac{\partial x_p}{\partial \theta_p} \\
        \frac{\partial y_p}{\partial \tau_p} & \frac{\partial y_p}{\partial \theta_p}
    \end{bmatrix}.
\end{equation}


\section{Two-stage Method for Ambiguity Problem Management} \label{sec:two_stage}


The advantage of using huge bandwidths at \ac{THz} is beneficial for both the communication and the sensing functionalities of the \ac{OFDM}-based \ac{ISAC} system.
From a communication viewpoint, wider bandwidths increase channel capacity and enhance robustness by exploiting multipath propagation. In contrast, for what concerns the sensing part, larger bandwidths enable higher range resolution, making it possible to distinguish targets that may be very close to each other. In fact, an \ac{OFDM} system with a bandwidth of the order of GHz can achieve cm-level range resolution, according to $r_\mathrm{res}=c/B$, for a bistatic configuration. 
Moreover, a larger bandwidth also increases performance in terms of \ac{RMSE} on position estimation, with a lower \ac{PEB}. Increasing the subcarrier spacing $\Delta f$ seems a natural choice to increase the bandwidth without compromising system complexity, especially when a \ac{ML} estimation approach is considered.

It's important to note that while increasing $\Delta f$ is feasible at THz due to the coherence band being of the order of hundreds of MHz \cite{Terahertz}, it can lead to an ambiguity problem in target position estimation. This problem persists even in relatively small areas, as the maximum unambiguous range, in these conditions, is less than the maximum distance that can be estimated in the area of interest. For instance, the maximum unambiguous bistatic range in an OFDM-based radar is $r_\mathrm{unamb} = c/\Delta f$. To solve this problem, which can cause the impossibility of correctly estimating the target position, a two-stage \ac{ML} estimation algorithm is proposed in this work. 
The approach consists of a novel two-stage transmission scheme and a two-stage \ac{ML} estimation approach at the receiver. In the first phase, a coarse target estimation exploits a transmitting signal that uses a sub-band with narrow subcarrier spacing to avoid the ambiguity problem. Then, in the second phase, the transmission exploits the full system bandwidth, and at the receiver, a refined super-resolution estimation is performed around the coarse estimate.

\begin{figure}[t]
    \centering
    \includegraphics[width=\columnwidth]{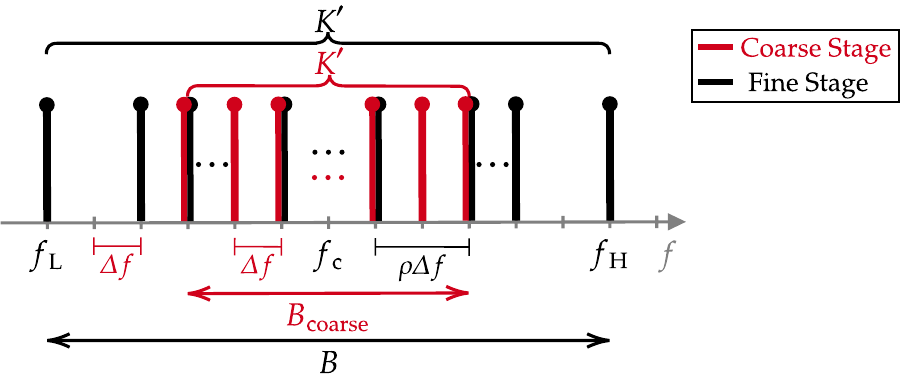}
    \caption{Illustration of the proposed two-stage method with $\rho=2$.}
    \label{fig:two_stage}
\end{figure}

The considered \ac{OFDM} \ac{ISAC} system operates over a large bandwidth $B$, of the order of GHz, with a large number of equispaced subcarriers $K$ and subcarrier spacing $\Delta f$, in the order of few MHz. However, such a system manages the complexity using a relatively small number $K'$ of active subcarriers out of the $K$ available, as shown in Fig.~\ref{fig:two_stage}. The remaining subcarriers can be used for communication purposes. The two phases are summarized below.
\begin{itemize}
    \item \emph{\textbf{Coarse target estimation}}: Localization of target $p$ with coarse resolution over a search area $A$ by using a sub-band $B_\mathrm{coarse}$ of the system bandwidth $B$ around the carrier frequency $f_\mathrm{c}$, with $K'$ contiguous active subcarriers within $B_\mathrm{coarse}$ and considering a subcarrier spacing $\Delta f$. The \ac{ML} target position estimation according to \eqref{eqn:ML_param} is performed without range ambiguity by properly designing $B_\mathrm{coarse}$ such that $B_\mathrm{coarse}\le(cK)/r_\mathrm{unamb}$, where $r_\mathrm{unamb}$ is the maximum unambiguous bistatic range compatible with the search area $A$.
    \item \emph{\textbf{Fine target estimation}}: Localization of target $p$ with refined resolution and without range ambiguity around the target position estimated in the coarse stage. Here, the system uses the same number $K'$ of properly equispaced active subcarriers out of the $K$ over the overall bandwidth $B$, with $\rho \Delta f$ subcarrier spacing, where $\rho \triangleq B/B_\mathrm{coarse}$. The search interval\footnote{Note that, the search interval width should be properly set, e.g., by considering the \ac{PEB} of the coarse stage, computed according to Section~\ref{sec:PEB}.} is centered in the target position estimated in the previous stage, and the width of the bistatic range search interval must be less than double the maximum unambiguous range, by considering the system working on the overall bandwidth $B$, i.e., $<2r_\mathrm{unamb}~=~(2cK)/B$. Again, the \ac{ML} approach in \eqref{eqn:ML_param} is here used for target position estimation.
\end{itemize}

This way, the proposed system can exploit an ultra-wide bandwidth at \ac{THz} to achieve highly precise and unambiguous range localization while maintaining low system complexity. 


\begin{figure*}[t]
    \centering
    \begin{subfigure}[b]{0.48\textwidth}
        \centering
        \includegraphics[width=\textwidth]{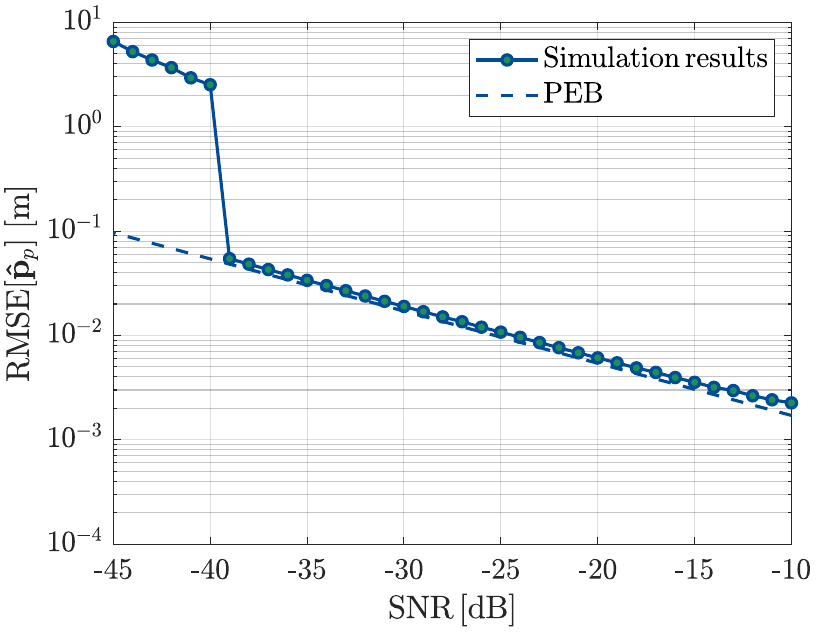}
        \caption{RMSE of target $p$ position estimation and PEB vs SNR}
        \label{fig:RMSE_vs_SNR}
    \end{subfigure}
    \hspace{0.5cm}
    \begin{subfigure}[b]{0.48\textwidth}
        \centering
        \includegraphics[width=\textwidth]{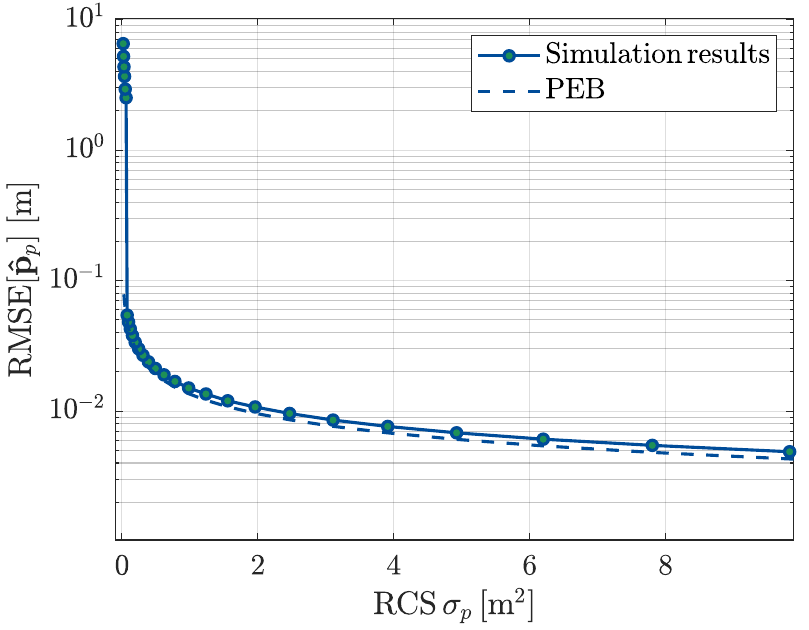}
        \caption{RMSE of target $p$ position estimation and PEB vs RCS $\sigma_p$}
        \label{fig:RMSE_vs_RCS}
    \end{subfigure}
    \caption{RMSE compared to the PEB related to position estimation $\mathbf{\hat{p}}_p$ of target $p$ as a function of \ac{SNR}~(a) and \ac{RCS} $\sigma_p$~(b), considering $r_{\mathrm{t},p}=r_{p,\mathrm{r}}=7.9\,$m, $\phi_p=18.4^\circ$, $\theta_p=-26.5^\circ$, and $|\mathbf{v}_p|=0\,$km/h as target parameters.}
    \label{fig:RMSE}
\end{figure*}

\section{Numerical Results} \label{sec:num_Res}
Numerical simulations are performed to evaluate the sensing performance of the proposed two-stage ultra-wideband system at \ac{THz} frequencies. In particular, we consider $f_\mathrm{c}=0.3\,$\ac{THz}, $B=2\,$GHz, $B_{\mathrm{coarse}}=400\,$MHz, $\Delta f=6.25\,$MHz, $K=320$, $K'=64$, $M=50$, $N_\mathrm{t} = N_\mathrm{r}=64$, $P_\mathrm{t}G^a_\mathrm{t} = 1\,$W, $G_\mathrm{r}=1$, $N_0=4\cdot10^{-20}\,$W/Hz, and $\rho=5$. 
The cyclic prefix duration has been set to $T_\mathrm{cp}=19.5\,$ns, according to the channel delay spread, here approximated as $T_\mathrm{cp}=(r_\mathrm{bis,max}-L)/c$, where $r_\mathrm{bis,max}$ is the maximum bistatic range within the considered squared area $A$ of $10\times10\,$~m$^2$. In addition, \ac{QPSK} constellation, and a scenario as the one described in Section~\ref{sec:system_model} are considered. The target $p$ is assumed to be stationary (i.e., $|\mathbf{v}_p|=0$), inside the designated area and located at a distance of $7.9\,$m from both \ac{Tx} and \ac{Rx} units, with $\phi_p=18.4^\circ$ and $\theta_p=-26.5^\circ$ as \ac{AoD} and \ac{AoA}, respectively, with the \ac{Tx} and \ac{Rx} placed in the bottom left and top right corners of the area. 

The analysis is performed through curves of \ac{RMSE} of target position estimation $\mathbf{\hat{p}}_p$ and \ac{PEB} as a function of the \ac{SNR} and target \ac{RCS} $\sigma_p$. 
When the \ac{SNR} is varied, this is normalized as explained in Section~\ref{sec:rx_signal_snr}. Conversely, when $\sigma_p$ is varied the considered \ac{SNR} is given according to \eqref{eq:SNR}, as a function of $\sigma_p$, $\mathbf{p}$, and the system parameters above.

The obtained results are shown in Fig.~\ref{fig:RMSE}. In particular, in Fig.~\ref{fig:RMSE_vs_SNR}, it can be observed that the considered \ac{ISAC} system with the proposed two-stage \ac{ML} estimation method, exhibits good position estimation performance up to $\mathrm{SNR}=-39\,$dB, where the \ac{RMSE} curve related to the target position estimation $\mathbf{\hat{p}}_p$ nearly coincides with the \ac{PEB} curve. The possibility of correctly estimating the target in correspondence with extremely low values of \ac{SNR} is especially valuable at \ac{THz}, where the path loss significantly affects the performance. Furthermore, it is worth noting that such a system can reach a mm accuracy on the position estimation, without range ambiguities, for values of \ac{SNR} greater than $-10\,$dB, under the above simulation parameters and conditions. Note that the bistatic range resolution is $15\,$cm, thanks to the use of a $B=2\,$GHz bandwidth.
In Fig.~\ref{fig:RMSE_vs_RCS}, the \ac{RMSE} on position estimation $\mathbf{\hat{p}}_p$ is plotted as a function of the \ac{RCS} of the target. As it can be observed, the proposed two-stage \ac{ISAC} system can correctly estimate, with sub-cm accuracy, the position of the target with an \ac{RCS} approximately greater than $1\,$m$^2$ at a distance of $7.9\,$m from both the \ac{Tx} and \ac{Rx}, showing performance very close to the theoretical bound.


\section{Conclusion} \label{sec:conclusion}
This work introduced a two-stage ML-based estimation approach for managing range ambiguity in a bistatic MIMO-OFDM ISAC system at THz frequencies that use a GHz-wide bandwidth. Initially, a coarse target position is estimated using a small sub-band to avoid ambiguity within the monitored area. The estimation is then refined using the system's full bandwidth to enhance position accuracy and resolution. The approach maintains low computational complexity and uses a small fraction of the subcarriers to save resources for communication. Numerical results confirm that the method effectively resolves range ambiguity, achieving near-optimal RMSE performance even at very low SNR regimes.
\bibliographystyle{IEEEtran}
\bibliography{IEEEabrv,bibliography}
\end{document}

%% file: acronyms.tex
\begin{acronym}
\acro{AoA}{angle of arrival}
\acro{AoD}{angle of departure}
\acro{AWGN}{additive white Gaussian noise}
\acro{BS}{base station}
\acro{CP}{cyclic prefix}
\acro{CRLB}{Cramér-Rao lower bound}
\acro{FOV}{field of view}
\acro{FIM}{Fisher information matrix}
\acro{FFT}{fast Fourier transform}
\acro{GML}{generalized maximum likelihood}
\acro{ICI}{inter-carrier interference}
\acro{ISAC}{integrated sensing and communication}
\acro{ISI}{inter-symbol interference}
\acro{LoS}{line of sight}
\acro{MIMO}{multiple-input multiple-output}
\acro{ML}{maximum likelihood}
\acro{OFDM}{orthogonal frequency-division multiplexing}
\acro{PEB}{position error bound}
\acro{QPSK}{quadrature phase shift keying}
\acro{Rx}{receiver}
\acro{RCS}{radar cross section}
\acro{RMSE}{root mean square error}
\acro{SNR}{signal-to-noise ratio}
\acro{THz}{terahertz}
\acro{Tx}{transmitter}
\acro{UE}{user equipment}
\acro{ULA}{uniform linear array}
\acro{6G}{sixth generation}
\end{acronym}

%% file: main.bbl
\begin{thebibliography}{10}
\providecommand{\url}[1]{#1}
\csname url@samestyle\endcsname
\providecommand{\newblock}{\relax}
\providecommand{\bibinfo}[2]{#2}
\providecommand{\BIBentrySTDinterwordspacing}{\spaceskip=0pt\relax}
\providecommand{\BIBentryALTinterwordstretchfactor}{4}
\providecommand{\BIBentryALTinterwordspacing}{\spaceskip=\fontdimen2\font plus
\BIBentryALTinterwordstretchfactor\fontdimen3\font minus \fontdimen4\font\relax}
\providecommand{\BIBforeignlanguage}[2]{{%
\expandafter\ifx\csname l@#1\endcsname\relax
\typeout{** WARNING: IEEEtran.bst: No hyphenation pattern has been}%
\typeout{** loaded for the language `#1'. Using the pattern for}%
\typeout{** the default language instead.}%
\else
\language=\csname l@#1\endcsname
\fi
#2}}
\providecommand{\BIBdecl}{\relax}
\BIBdecl

\bibitem{Tong_Zhu_2021}
W.~Tong and P.~Zhu, \emph{6G: The Next Horizon: From Connected People and Things to Connected Intelligence}.\hskip 1em plus 0.5em minus 0.4em\relax Cambridge University Press, 2021.

\bibitem{THz_channel}
A.~Hrovat \emph{et~al.}, ``Integrated communications and sensing in terahertz band: A propagation channel perspective,'' \emph{J. Commun. Softw. Syst.}, vol.~20, no.~1, pp. 23--37, 1 2024.

\bibitem{THz_survey}
W.~Jiang \emph{et~al.}, ``Terahertz communications and sensing for 6g and beyond: A comprehensive review,'' \emph{{IEEE} Commun. Surveys Tuts.}, 2024.

\bibitem{Barneto2019}
C.~B. Barneto \emph{et~al.}, ``Full-duplex {OFDM} radar with {LTE} and {5G NR} waveforms: challenges, solutions, and measurements,'' \emph{{IEEE} Trans. Microw. Theory Techn.}, vol.~67, no.~10, pp. 4042--4054, Oct. 2019.

\bibitem{Braun}
M.~Braun, ``{OFDM} radar algorithms in mobile communication networks,'' Ph.D. dissertation, Karlsruhe Institute of Technology, 2014.

\bibitem{pucci2022system}
L.~Pucci, E.~Paolini, and A.~Giorgetti, ``System-level analysis of joint sensing and communication based on {5G} new radio,'' \emph{{IEEE} J. Sel. Areas Commun.}, vol.~40, no.~7, pp. 2043--2055, July 2022.

\bibitem{Pucci_Bistatic}
L.~Pucci, E.~Matricardi, E.~Paolini, W.~Xu, and A.~Giorgetti, ``Performance analysis of a bistatic joint sensing and communication system,'' in \emph{IEEE Int. Conf. on Commun. Work.}, Seoul, Republic of Korea, May 2022, pp. 73--78.

\bibitem{Beam_THz}
F.~Gao, L.~Xu, and S.~Ma, ``Integrated sensing and communications with joint beam-squint and beam-split for mmwave/{THz} massive {MIMO},'' \emph{{IEEE} Trans. Commun.}, vol.~71, no.~5, pp. 2963--2976, 2023.

\bibitem{THz_ISAC}
C.~Han, Y.~Wu, Z.~Chen, Y.~Chen, and G.~Wang, ``{THz} {ISAC}: A physical-layer perspective of terahertz integrated sensing and communication,'' \emph{{IEEE} Commun. Mag.}, vol.~62, no.~2, pp. 102--108, 2024.

\bibitem{dehkordi2023multistatic}
\BIBentryALTinterwordspacing
S.~K. Dehkordi, L.~Pucci, P.~Jung, A.~Giorgetti, E.~Paolini, and G.~Caire, ``Multi-static parameter estimation in the near/far field beam space for integrated sensing and communication applications,'' 2023. [Online]. Available: \url{https://arxiv.org/abs/2309.14778}
\BIBentrySTDinterwordspacing

\bibitem{Rife1974}
D.~Rife and R.~Boorstyn, ``Single tone parameter estimation from discrete-time observations,'' \emph{{IEEE} Trans. Inf. Theory}, vol.~20, no.~5, pp. 591--598, 1974.

\bibitem{BacPucPaoGio:C23}
T.~Bacchielli, L.~Pucci, E.~Paolini, and A.~Giorgetti, ``Performance analysis of a low-complexity {OTFS} integrated sensing and communication system,'' in \emph{IEEE Vehicular Technology Conference}, Hong Kong, Hong Kong, Oct. 2023.

\bibitem{jourdan2008position}
D.~B. Jourdan, D.~Dardari, and M.~Z. Win, ``Position error bound for {UWB} localization in dense cluttered environments,'' \emph{{IEEE} Trans. Aerosp. Electron. Syst.}, vol.~44, no.~2, pp. 613--628, 2008.

\bibitem{Terahertz}
I.~F. Akyildiz, C.~Han, Z.~Hu, S.~Nie, and J.~M. Jornet, ``Terahertz band communication: An old problem revisited and research directions for the next decade,'' \emph{{IEEE} Trans. Commun.}, vol.~70, no.~6, pp. 4250--4285, 2022.

\end{thebibliography}
